\documentclass[twocolumn,showpacs,preprintnumbers,aps,eqlabels]{revtex4}

\usepackage{graphicx}
\usepackage{dcolumn}
\usepackage{bm}
 
\def\mb#1{{\boldsymbol #1}}
\def\bm{\boldsymbol}

\def\bea{\begin{eqnarray}}
\def\eea{\end{eqnarray}}

\def\ben{\begin{equation}}
\def\een{\end{equation}}

\input rotate

\begin{document}

\preprint{DIPC}

\title{An efficient method for the Quantum Monte Carlo evaluation of 
the static density-response function of a many-electron system}
\author{R. Gaudoin$^1$ and J. M. Pitarke$^{2,3}$}
\affiliation{
$^1$Donostia International Physics Center (DIPC),
Manuel de Lardizabal Pasealekua, E-20018 Donostia, Basque Country, Spain\\
$^2$CIC nanoGUNE Consolider, Tolosa Hiribidea 76, E-20018 Donostia, Basque
Country, Spain\\
$^2$Materia Kondentsatuaren Fisika Saila, UPV/EHU, and Centro F\'\i sica Materiales
CSIC-UPV/EHU, 644 Posta kutxatila, E-48080 Bilbo, Basque Country, Spain}

\date{\today}

\begin{abstract}
In a recent Letter we introduced Hellmann-Feynman operator sampling
in diffusion Monte Carlo calculations. Here we derive, by evaluating
the second derivative of the total energy, an efficient method for the
calculation of the static density-response function of a many-electron
system. Our analysis of the effect of the nodes suggests that
correlation is described correctly and we find that 
the effect of the nodes can be dealt with.
\end{abstract}

\maketitle

\section{Introduction}

Diffusion Monte Carlo (DMC) represents a powerful method for the
accurate computation of properties of molecules and
solids~\cite{qmc}. However, so far few attempts \cite{cep,bsa} have
been made to use DMC to calculate the static density-response
function~\cite{fetter}, which is a central quantity in the
analysis of many-electron systems and time-dependent density-functional 
theory~\cite{tddft}. One reason is the
technical difficulty inherent in the most straightforward method to do
so: For a given perturbing potential one calculates the total
energy at different strengths and numerically determines the second
derivative. This then gives a DMC estimate of the diagonal term of the
static response function $\chi$.  There are, however, several obvious
difficulties with this.  One needs one loop for various perturbation
strengths, another loop for each $k$ that one wishes to sample, and if
one wants the off-diagonal terms a third loop for the $k'$. Inside
each of these loops sits an entire wavefunction reoptimization cycle
and a complete DMC run. The perturbations must be small enough not to
change the wavefunction qualitatively and large enough to allow for
sensible numerical derivatives.

In a recent Letter~\cite{hfs}, we showed how ``applying'' the
Hellmann-Feynman~\cite{hf} (HF) derivative to the DMC algorithm leads to 
a new algorithm, Hellmann-Feynman sampling (HFS),
 that correctly samples the first derivative
of the energy, i.e., an expectation value of an operator. 
HFS works because DMC yields the correct total
energy for nodes defined by the trial wavefunction. 
For technical reasons the operators sampled must be
diagonal in real space. Extending the analysis to the second
derivative yields a DMC algorithm for the fixed-node (fn) static
density-response function. Note that even for a trial wavefunction
with correct nodes the fixed-node density response is not the
exact value as the real response includes effects from the change of
the nodes. However, comparison with Ref.~\cite{cep} where 
the nodal variation of an underlying Kohn-Sham (KS) system is 
implicitly used, shows that
these effect can be 
accounted for by generalizing the RPA analysis~\cite{fnrpa} to fn systems.
The resulting method can be performed within a single DMC
run, and in the case of inhomogeneous systems
 can produce off-diagonal elements of $\chi$ as easily as the
diagonal terms.

The present paper is organized as follows. After a brief
recapitulation of HF sampling, we derive formulae for the DMC
sampling of $\chi$ along the same line. We then briefly discuss
technical aspects (convergence with respect to population size, time
step, etc.).  Finally, we look at the density response of the
interacting and non-interacting uniform electron gas, analyze the
effects of the nodes, and compare our results with the
literature. Our method should also enable DMC calculations of the
static-response function of real solids, never done before.  We use
atomic units throughout.
 
\section{Hellmann-Feynman sampling and the density response function}
\subsection{Application to the second derivative of the energy}
  Fixed-node DMC yields by
construction the normalized expectation value $\langle
\hat{O}\rangle_{DMC}= \langle \Psi_T| \hat{O}|\Psi^{fn}_0
\rangle/\langle \Psi_T|\Psi^{fn}_0\rangle$, where $\Psi^{fn}_0$ is the
ground-state wavefunction constrained by the nodes of the
Fermionic many-body trial wavefunction $\Psi_T$;
HFS correctly calculates  $\langle \Psi^{fn}_0| \hat{O}|\Psi^{fn}_0 
\rangle/\langle \Psi^{fn}_0|\Psi^{fn}_0\rangle$ while maintaining the basic DMC 
algorithm that samples $\Psi_T\Psi_0^{fn}$. This is because the 
total energy is evaluated correctly within standard DMC, and
crucially operator
expectation values can be cast as HF derivatives of the total energy.
Keeping in mind that ultimately the DMC algorithm is nothing but a
large sum that yields the total energy, we see that the HF derivative
can be applied to the algorithm itself!  One advantage
over numerical derivatives is that the resulting formula can handle
several operators simultaneously in a single DMC run, and maintaining
orbital occupancy for perturbed Hamiltonians ceases to be a problem.
The DMC algorithm only involves numbers, so non-commutability of
operators is no problem. 
Writing down the DMC algorithm as a mathematical formula and
applying the HF derivative to it yields an object that when sampled
using standard DMC produces the {\it exact} operator expectation
value. We find that given a Hamiltonian
$\hat{H}(\alpha)=\hat{H}+\alpha \hat{O}$, evaluating the growth
estimator of the energy $E^{GR}$ at time step $i$ to first order in $\alpha$ 
yields a growth
estimator that samples the operator $\hat O$. Similarly the direct estimator
$E$ of the energy yields another estimator. These are  
Eqs.~ (8)  and (9) of Ref.~\cite{hfs}:
\bea
\label{def_de0}
O^{GR}_i&=&\left.\frac{\partial E^{GR}_i(\alpha)}{\partial\alpha}\right|_{\alpha=0}=
\overline{ X_{i}}.
\\
\label{def_de1}
O^{E}_i&=&\left.\frac{\partial
  E_i(\alpha)}{\partial\alpha}\right|_{\alpha=0}= \overline{ O^L_{i} }
-t\left( \overline{ E^L_{i} X_{i} } -\overline{ E^L_{i}} \cdot
\overline{ X_{i}} \right).  
\eea 
where the bar refers to the DMC
sampling at time step $i$: $\overline{
  X_{i}}= \sum_j^{N_w} \omega_{i,j} X_{i,j} $. 
$ \omega_{i,j}$ is the
total weight of walker $j$,
$X_{i,j}=\frac{1}{i}\sum_{k=1}^{i} O^L_{k,j}$, and $O^L_{k,j}$ is
$\frac{\hat{O}\Psi_T}{\Psi_T}$ evaluated for walker $j$ at time step $k$.
Now we assume two perturbations of the form $\alpha\hat{O}_A$ and
$\beta\hat{O}_B$, and following Ref.~\cite{hfs} we obtain 
growth and direct estimators of the response function 
$\chi_{AB}=\frac{\partial \hat{O}_A}{\partial \beta}=\frac{\partial \hat{O}_B}{\partial \alpha}$ 
from the second derivative of the growth and direct estimators of the energy:
\bea
\label{chi_gr}
\chi^{GR}_{AB}(i)&=&
-t[\overline{X^A_{i}X^B_{i}}-\overline{X^A_{i}}\cdot\overline{X^B_{i}}]
\\
\nonumber
\chi^E_{AB}(i)&=&
-t[\overline{X^A_{i}O^B_{i}}-\overline{X^A_{i}}\cdot\overline{O^B_{i}}
+\overline{O^A_{i}X^B_{i}}-\overline{O^A_{i}}\cdot\overline{X^B_{i}}]
\\
\label{chi_dir}
&&
+t^2[\overline{
(E^L_{i}-\overline{E^L_{i}})(X^A_{i}-\overline{X^A_{i}})
(X^B_{i}-\overline{X^B_{i}})}
]
\, .
\eea
Any number of operators $\hat{O}^A$ and $\hat{O}^B$ can be
sampled in parallel within a single DMC run. From now on we 
use the Fourier components of
the density $\sin {\mb k}{\bm x}$ and $\cos {\mb k}{\bm x}$. Note also
that, as in the case of the first derivative discussed in Ref.~\cite{hfs}, 
the growth estimator at $i$
is already an averaged quantity.  This property makes it an
attractive choice for a DMC calculation.  Equation~(\ref{chi_dir}) is not
only a more complicated formula than Eq.~(\ref{chi_gr}), it also has
to be summed at each time step $i$. If we wish to sample many
components of $\chi$ at the same time, a large 
array with a size quadratic in the
number of components of $\chi$ in Eq. (\ref{chi_dir}) has to be
generated and dealt with at each time step.  By contrast,
Eq. (\ref{chi_gr}) only has to be built at whatever time step one wishes 
to calculate $\chi$. This means that at
each time step one only has to maintain the $X^{A/B}_{i}$, which is
less memory intensive and much faster to compute. Hence we shall only
use the growth estimator Eq.~(\ref{chi_gr}).

\subsection{Computational implementation}  

The growth estimator has the advantage that for each time step and
walker we only need to deal with simple sampling of $N_k$ variables
for the components of the density.  The entire density-response
function, including off-diagonal elements, can then be calculated as a
correlation function~\cite{vbrik} of these variables at the end of the
run saving computer time and memory.  As in HFS, we find that noise
rises as the sampling progresses, thus limiting the statistical error
of the final result even if the sampling is continued indefinitely. So
to reduce statistical noise, we increase the number of walkers
instead. This has the additional advantage of reducing any population
bias. We converged this by looking at population sizes of $200$,
$1000$, $5000$, and $50000$ walkers. We used the latter for the main
results shown here (at $N=114$ electrons). Looking at different time
steps, we found that too large a time step shows up as a levelling off
of $\chi$ at a finite value at larger $k$ instead of showing the
correct $1/k^2$ behavior. Interestingly, even at $\delta t=0.1$ the
calculated $\chi$ remained unbiased up to and well beyond $k=5k_F$.  A
large time step is desirable as equilibration will be faster. Here we
use $\delta t = 0.01$.  To monitor equilibration we artificially
extract a value $\tilde{\chi}_i$ that when summed over all time steps
gives the growth estimator $\chi(N)$ at the time $t=N\Delta t$ of
sampling: From 
\ben
\frac{1}{N-1} \sum_i^{N-1}\tilde{\chi}(i)=\chi(N-1)
\een
and 
\ben\frac{1}{N} \sum_i^{N}\tilde{\chi}(i)=\chi(N)
\een
 it follows that
\ben
\tilde{\chi}(N)=N\chi_N-(N-1)\chi(N-1)\ .
\een

 Following, e.g. three
typical $k$'s as the sampling progresses we found that $\tilde{\chi}$
converges exponentially. A quick run using few walkers in a smaller
system can then be used to roughly estimate the convergence time which
one then uses in an actual run. Convergence can be improved by setting
up the sampling such that one ignores the implied $\tilde{\chi}$
during equilibration. In order to do that it is not necessary to
actually reverse engineer these $\tilde{\chi}$ for each element of the
density response function. Once one has decided on an equilibration
time, $N_{eq}$ the desired result is
\ben
\frac{1}{N-N_{eq}}\sum_{i=N_{eq}+1}^{N} \tilde{\chi}(i)=
\frac{1}{N-N_{eq}}[N\chi(N)-N_{eq}\chi(N_{eq})]
\een
It is thus sufficient to
perform the costly sampling of the full growth estimator only twice
during the run, once after equilibration, and once at the end of the
run.  Doing so, efficiently removes much of a $1/N$-like term without
incurring much extra computation. We found that equilibration was
essentially independent of $\delta t$ and seemed to depend only weakly
on the population size.  We note that while in this paper we only
calculate the diagonal terms of $\chi$, sampling the full
density-response function, including all the off-diagonal terms, is
not more difficult: All that is needed is the evaluation of the
correlation function of $X_k$s at different $k$ and $k'$.

\section{Results}
\subsection{The system}
 In order to demonstrate our method we calculated the
diagonal terms of the static density response function of an
unpolarized free electron gas for electron-density parameter 
$r_s= 2$, $5$, and $10$ and corresponding density $n_0$.  We set
up the DMC calculation using $114$ electrons in a simple-cubic super
cell.  We also looked at $fcc$ unit cells and smaller systems with
$66$ electrons, however, we found no qualitative difference. 
In all, we calculated the
density response at all $119$ independent $k$-vectors between $k=0$
and $k=5k_F$. Our DMC calculations employed trial wave functions
$\Psi_T$ of the Slater-Jastrow type with a standard correlation
term. Prior to the DMC run $\Psi_T$ was optimized in a variance
minimization run. We used the CASINO~\cite{casino} code for all our
computations.

\begin{figure}
\centering
\includegraphics[width=0.48\textwidth]{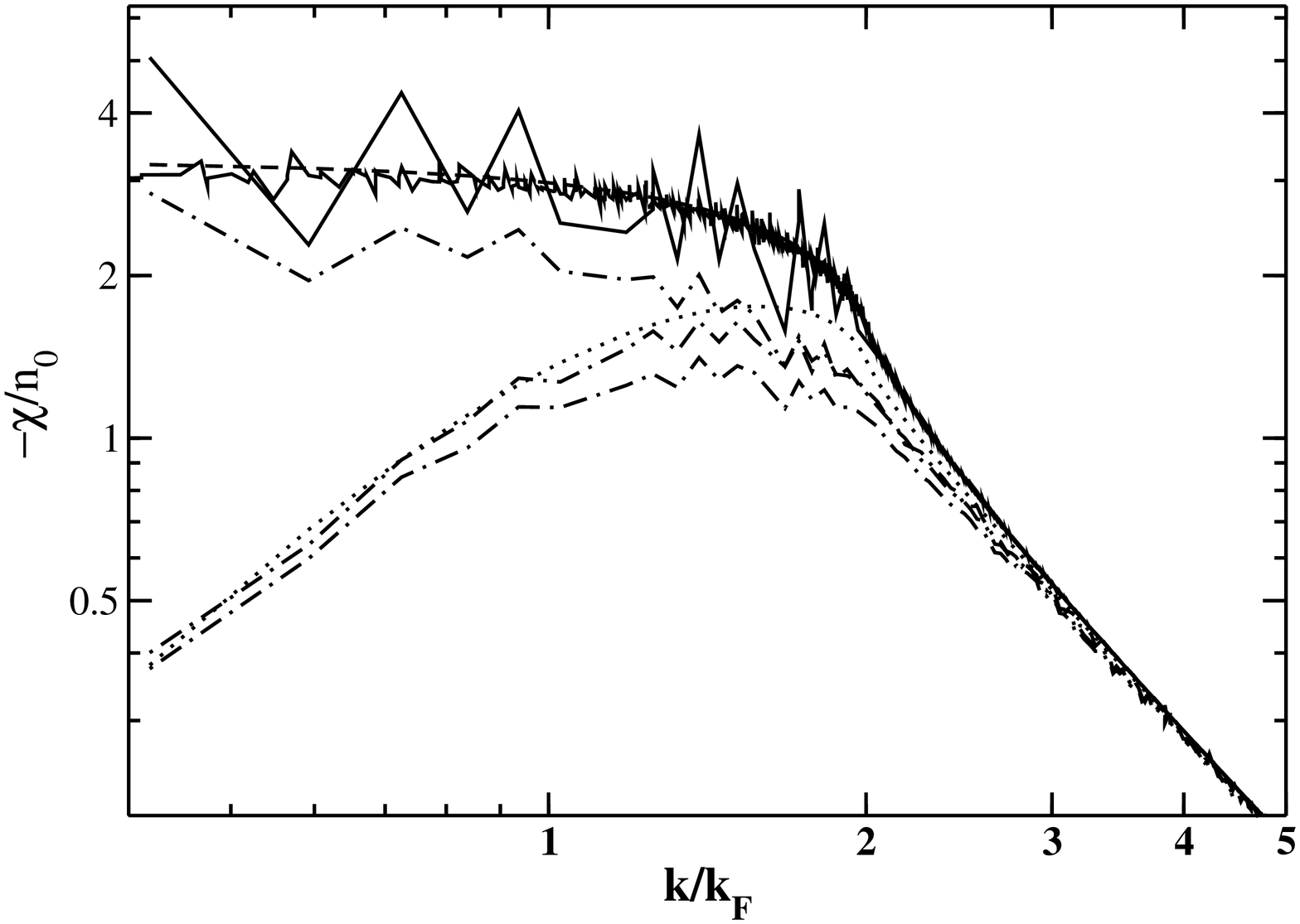}
\caption{ The dashed line shows the density response function $\chi_0$
  of an infinitely large unpolarized non-interacting homogeneous
  electron gas (Lindhard function) at $r_s=2$. The black lines close
  to the Lindhard function show the exact finite-size Lindhard
  function $\chi_0^{fs}$with $114$ and $4218$ electrons, the latter
  following the Lindhard function closer.  The dotted line shows the
  RPA response function $\chi_{RPA}$, and the three remaining dot-dashed
  lines show (from top to bottom) the fixed-node Lindhard function
  $\chi^{fn}_0$ at $114$ electrons, the corresponding fixed-node
  density response of an interacting system $\chi^{fn}$, and the
  fixed-node RPA $\chi_{RPA}^{fn}$. The ``wiggles'' are not noise as
  they correspond to the shell structure seen in the exact
  non-interacting finite-size $\chi_0^{fs}$}
\label{chi1}
\end{figure}

\subsection{The density-response function}
 In general, the response function is given by
\ben
\chi_{AB}=\sum_{i=1}^{\infty}2\Re \frac{\langle
0|\hat{O}_A|i\rangle\langle i|\hat{O}_B|0\rangle} {E_0-E_i},
\label{fnx}
\een 
where the sum runs over all excited states of the many-electron
system. fn DMC yields the ground-state energy for nodes
given by the trial wave function. Therefore the second derivative
yields the fixed-node response $\chi^{fn}$ of a system for which the
nodes are the same for all perturbing potentials. Since in the case of
a fixed-node system the sum entering Eq.~(\ref{fnx}) runs over a set
of fixed-node excited states that differ from the actual excited
states of the many-electron system, the fixed-node and non fixed-node
non-interacting density-response function
 differ considerably (Fig.~\ref{chi1}). Another interesting observation is the
shell structure exhibited by all finite-size (fs) results. In order to
visualize both the fixed-node error and the finite-size effects, we
have plotted in Fig.~\ref{chi1} the following calculations: the
static density-response function of (i) an infinitely large
unpolarized non-interacting free electron gas, the well-known
Lindhard function~\cite{pines} $\chi_0$ (dashed line), (ii) $\chi_0^{fs}$ of a
non-interacting unpolarized system of 114 and 4218 electrons (solid lines), 
(iii) an infinitely large
unpolarized interacting free electron gas in the random-phase
approximation (RPA), $\chi_{RPA}$ (dotted line), 
(iv) a
finite unpolarized system of 114 non-interacting electrons within the
fixed-node approximation giving $\chi^{fn}_0$ (top dot-dashed line), 
(v) a finite unpolarized system of
114 interacting electrons within our fixed-node DMC scheme 
(middle dot-dashed line), and (vi) a finite 
unpolarized system of 114 interacting electrons in the fn RPA
$\chi_{RPA}^{fn}$ (dot-dashed line at the bottom,
see below for details). 
We see that the finite-size shell structure is not
negligible even for a system of 4218 electrons. Nevertheless, 
the exact non-interacting density-response function nicely
reproduces  the well-known Lindhard
function especially for $k/kf>2$. 

\begin{figure}
\centering
\includegraphics[width=0.48\textwidth]{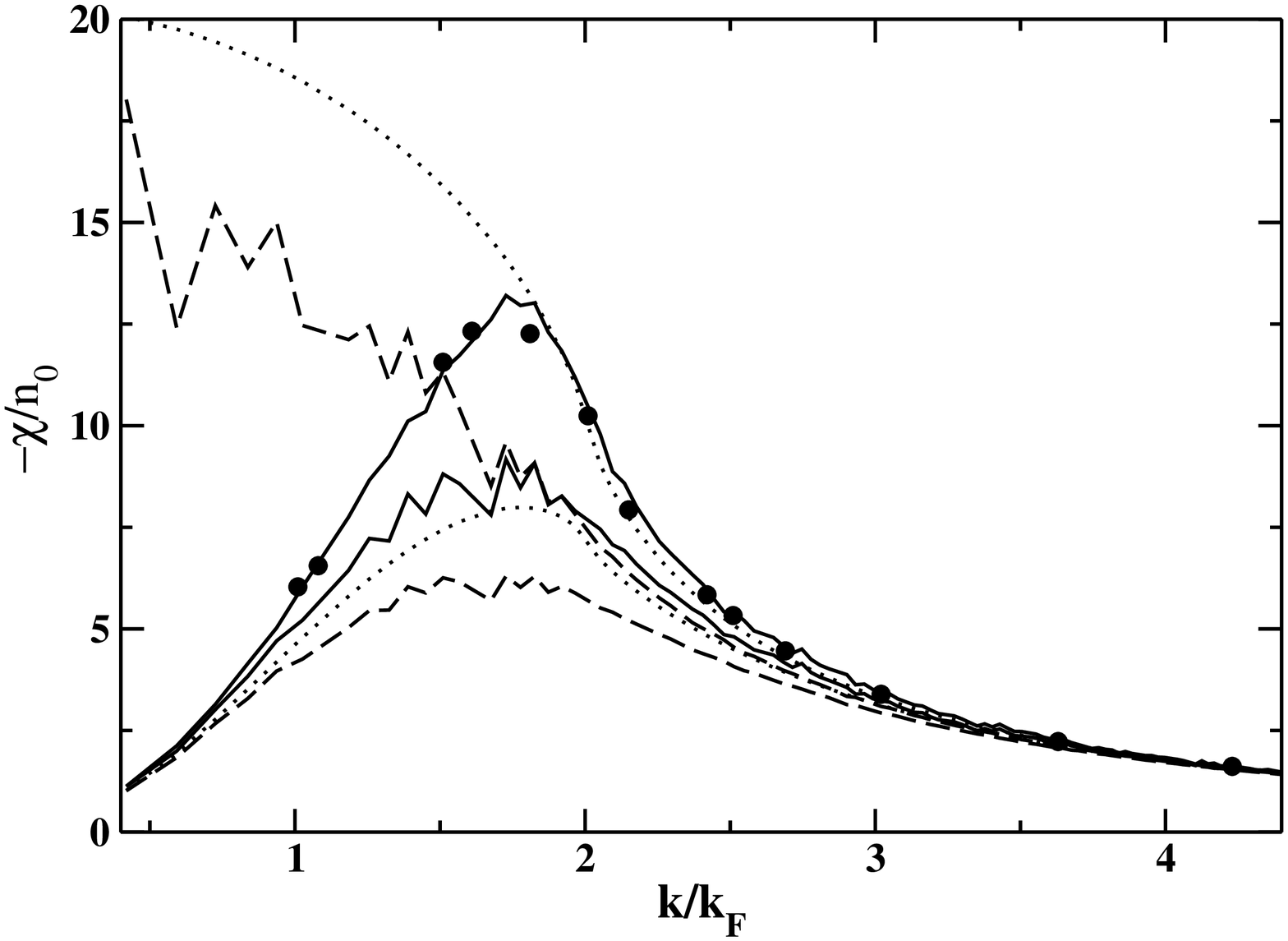}
\caption{ The density response at $r_s=5$ for an unpolarized $114$
  particle homogeneous electron gas.  The dotted lines show the
  Lindhard function $\chi_0$ (top) and $\chi_{RPA}$ (bottom).  The
  dashed lines are $\chi_0^{fn}$ (top) and $\chi_{RPA}^{fn}$
  (bottom). Finally, the solid lines show the fixed-node $\chi^{fn}$
  (bottom) and extrapolated interacting $\chi$. The dots correspond to
  the results in Ref.~\cite{cep}. Our data reaches to smaller $k$
  values as our system is larger ($114$ vs. up to $66$ electrons).}
\label{chis}
\end{figure}

\subsection{Discussion}
\begin{figure}
\centering
\includegraphics[width=0.48\textwidth]{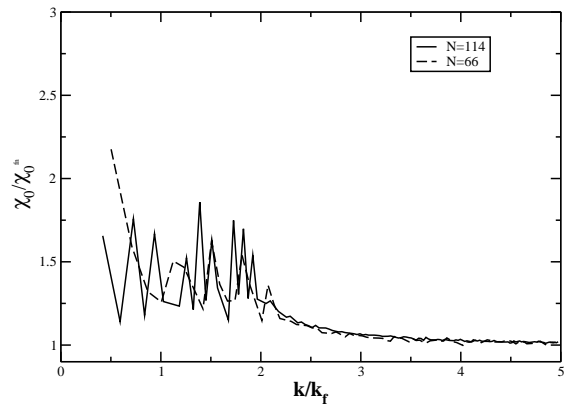}
\caption{ The ratio $\frac{\chi_0}{\chi_0^{fn}}$ at $r_s=5$ for
  $N=66$ and $N=114$ electrons. Due to scaling this graph is
  independent of the value of $r_s$, except for noise. Note the
  pronounced shell structure for $k<2k_f$.
}
\label{uninode}
\end{figure}

\begin{figure}
\centering
\includegraphics[width=0.48\textwidth]{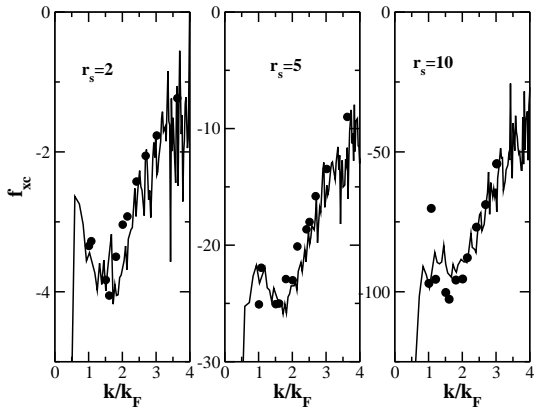}
\caption{ $f_{xc}^{fn}$ at $r_s=2, 5, 10$. The value at
  $k=0.4189 k_F$ clearly is an outlier. Also, the noise
increases as $k$ grows. Interestingly, there is a slight dip in $f_{xc}^{fn}$ 
for 
$k/k_F<2$ as demonstrated in Ref.~\cite{fxcref}. 
The dots correspond to  the data in  Ref.~\cite{cep}. 
 }
\label{fxc_5}
\end{figure}

\begin{figure}
\centering
\includegraphics[width=0.48\textwidth]{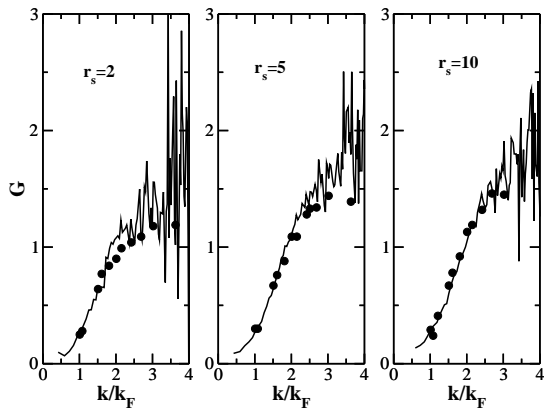}
\caption{ As figure \ref{fxc_5}, but showing the local field factor.}
\label{g_5}
\end{figure}

\begin{figure}
\centering
\includegraphics[width=0.48\textwidth]{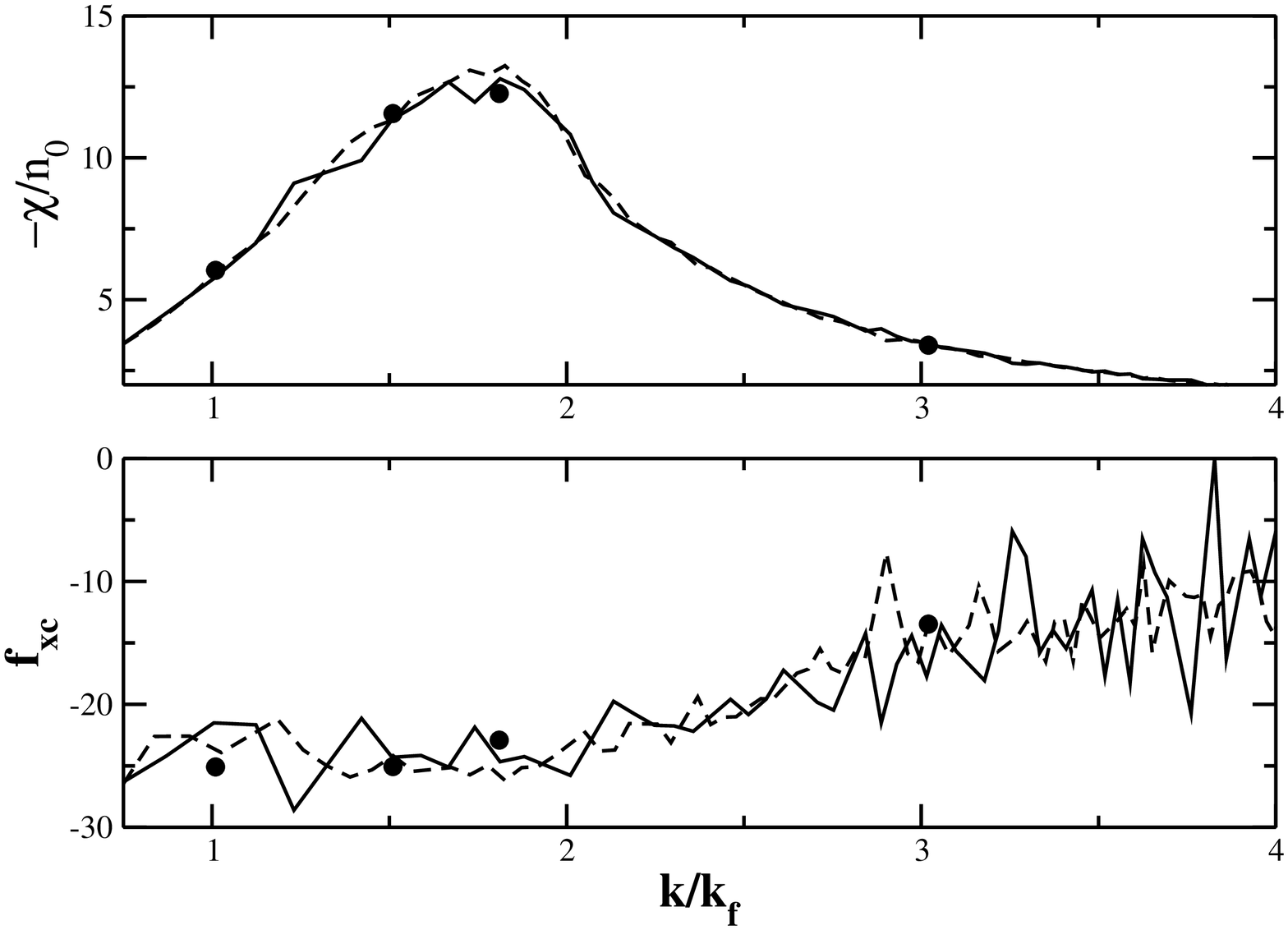}
\caption{A comparison of our results for $f_{xc}$ at $r_s=5$ and the finite-size
  corrected $\chi$ with the data in Ref.~\cite{cep} where comparison
  is possible, i.e. at $N=66$ electrons. All our other
  data uses a larger simulation cell with $114$ electrons and more
  configurations in the DMC run. The solid dots are the four data points
  for $N=66$ electrons of Ref.~\cite{cep} and the solid line
  corresponds to our data (just under 100 data points in the shown
  region). The dashed line shows our data for the $N=114$ electron
  system. As suggested in Ref.~\cite{cep} there is no qualitative
  difference between $N=66$ and $N=114$.
Note that for $N=66$ we use fewer walkers than for our
main results at $N=114$.
 }
\label{ccomp}
\end{figure}

 Since the fn Lindhard function
$\chi^{fn}_0$ (top dashed line in Fig.~\ref{chis}) is smaller than the real
 Lindhard function $\chi_0$ (dotted line at the top of  Fig.~\ref{chis}) 
the fn interacting $\chi^{fn}$  is
also too small (lower solid line in Fig.~\ref{chis} compared 
to the dots showing the relevant data of Ref.~\cite{cep}). 
However, assuming that the effect of the fixed-node nature
of the calculation is the same for the interacting and non-interacting
case, we should still be able to extract meaningful correlation data
and reverse the effects of the fn approximation ~\cite{fnrpa}. 
Let us start with the definition for the exchange correlation (xc)
kernel $f_{xc}$ and the local field factor $G$ 
\bea \nonumber
-f_{xc}(k)&=&v_C(k)G(k)= \frac{1}{\chi(k)}-\frac{1}{\chi_{RPA}(k)} \\
\label{fxc}
&=& \frac{1}{\chi(k)}-\frac{1}{\chi_{0}(k)}+v_C(k) \eea where
$v_C(k)=\frac{4\pi}{k^2}$.  In practise, we do not have access to
$\chi(k)$ 
but only to its finite-size equivalent
in the fixed-node approximation.  Moroni and coworkers ~\cite{cep},
who include the nodal variation at a Kohn-Sham level argue that that
while the density response contains finite size effects, $f_{xc}$ is
less afflicted by these.  Hence, they extract $f_{xc}$ and add that
back on to the non fixed-node infinite-cell Lindhard function to 
correct for finite size effects,
thus eliminating the shell structure. There is no reason to expect the
nodal variation of the KS nodes to correctly describe the nodal
variation of the fully interacting system with respect to $\chi$: The
KS nodes and the true many-body nodes are unrelated. Furthermore,
different QMC systems at different numbers of electrons $N$ also
correspond to distinct nodes, but the data for $f_{xc}$ (e.g.
Ref.~\cite{cep} or  Fig.~\ref{ccomp})
 for different values of $N$  is mutually compatible.
Finally,
the effect of the nodal variation on $\chi_0$ seems universal,
i.e. independent of $N$ except for shell effects (see Fig.~\ref{uninode}).
It therefore seems reasonable to assume that $f_{xc}$ is independent of
nodal effects and can thus be used to correct for wrong or absent nodal
variation.  Hence, by using the fixed-node quantities in
Eq. (\ref{fxc}), implicitly defining a fixed-node $\chi_{RPA}^{fn}$,
we can derive a fn $f_{xc}^{fn}$ and $G^{fn}$ (Figures~ \ref{fxc_5}
and \ref{g_5}). These are remarkably similar to the data in
Ref.~\cite{cep}. In fact, $f_{xc}^{fn}$ even has a slight dip as
suggested in Ref.~\cite{fxcref} which however is not really visible in
Ref.~\cite{cep}.  This is encouraging and indeed we can use our data
for $G^{fn}$ in conjunction with the real $\chi_{RPA}$ in
Eq. (\ref{fxc}) to estimate the non-fn interacting $\chi$. The result
can be seen in Fig.~ \ref{chis}: All the fn quantities are too small
compared to their non-fn counterparts. However our extrapolated date
(solid line at the top) is very close the extrapolated data of
Ref.~\cite{cep} (dots).
%
%

\begin{figure}
\centering
\includegraphics[width=0.48\textwidth]{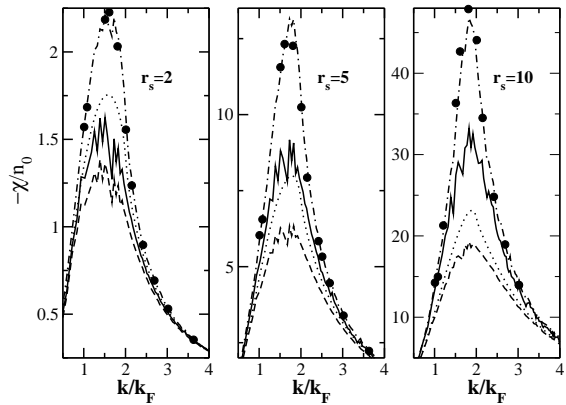}
\caption{
RPA (dotted), fn RPA (dashed), fn DMC (solid), and extrapolated (dot-dashed) 
density response function
calculated using a $114$ particle homogeneous electron gas at $r_s=$ 
$2$,$5$, and $10$. The dots correspond to the results in Ref.~\cite{cep}.
Note that for $r_s=2$ the uncorrected DMC data is lower still than standard RPA.
}
\label{chi2}
\end{figure}

For completeness sake Fig.~ \ref{chi2} shows details of the
extrapolated $\chi$ at $r_s=2$,$5$, and $10$.  Also, in
Fig.~\ref{ccomp} we show a direct comparison between our results and
Ref.~\cite{cep} where it is possible, i.e. at $N=66$ electrons in
addition to our results for $N=114$ electrons confirming that 
all our data is compatible with Ref.~\cite{cep}.  Except for noise there
is no significant difference between data at different $N$,
corroborating the assumption that $f_{xc}$ is independent of
finite-size effects.

In general, our results nicely follow the data in Ref.~\cite{cep},
who take into account the change in the nodes at a Kohn-Sham
level, 
whereas our calculations do not take into account any nodal effect on
xc quantities. The fact that the methods yield consistent results for
$f_{xc}$ suggest that assuming $f_{xc}$ to be free from nodal effects
is justified and that in either case the resulting data is an accurate
description of systems with the full interacting nodal variation.

\section{ Concluding remarks} We have generalized Ref.~\cite{hfs} to the
second derivative of the energy.  This yields a novel method and an efficient
algorithm to calculate the static response function within DMC.  Our
algorithm permits the computation of a large number of diagonal and
off-diagonal terms in a single DMC run without the need for numerical
derivatives or re-optimization. Noise can be efficiently controlled by
increasing the number of DMC walkers and we have found that we can use
large DMC time steps without introducing a bias, potentially speeding
up calculations greatly. The wavefunction nodes have a
strong effect on $\chi$, particularly for $k<3k_F$ and generalizing the
RPA analysis using $\chi_0^{fn}$ yields a fixed-node $\chi_{RPA}^{fn}$. 
Using this to extract
the xc contribution of $\chi$ we find that our 
method's results are broadly in line with
previous DMC calculations ~\cite{cep} which, however, are much more
cumbersome, yield potentially fewer data points, and are effectively
limited to diagonal terms only. 

This work has been supported by the Basque Unibertsitate eta Ikerketa
Saila and the Spanish Ministerio de Ciencia e Innovacion (Grants
No. FIS2006-01343 and CSD2006-53).  Computing facilities were provided
by the Donostia International Physics Center (DIPC) and the
SGI/IZO-SGIker UPV/EHU (supported by the National Program for the
Promotion of Human Resources within the National Plan of Scientific
Research, Development and Innovation-Fondo Social Europeo, MCyT and
the Basque Government).



\begin{thebibliography}{100}

\bibitem{qmc} W. M. C. Foulkes, L. Mitas, R. J. Needs, and
  G. Rajagopal, Rev.  Mod. Phys. {\bf 73}, 33 (2001).
\bibitem{cep} S. Moroni, D. M. Ceperley, and G. Senatore,
Phys. Rev. Lett. {\bf 75}, 689  (1995).

\bibitem{bsa} C Bowen, G. Sugiyama, and B. J. Alder, Phys. Rev. {\bf 50}, 14838 (1994).

\bibitem{fetter} A. L. Fetter and J. D. Walecka, Quantum Theory of
Many-Particle Systems, MgGraw-Hill, New York, 1971.

\bibitem{tddft}  E. Runge and E. K. U. Gross, Phys. Rev. Lett. 
{\bf 52}, 997 (1984).


\bibitem{hfs} R Gaudoin, J.M. Pitarke, Phys. Rev. Lett. {\bf 99}, 126406 (2007).

\bibitem{hf} R. P. Feynmann, Phys. Rev. {\bf 56}, 340 (1939).

\bibitem{fnrpa} Note that the analysis leading to RPA (e.g. Ref.~\cite{fetter})
  does not mention nodes and thus should apply equally to the fn
  system.


\bibitem{vbrik} Another method (J. Vrbik and S.M. Rothstein, J. Chem. 
Phys. {\bf 96}, 2071 1992.) to calculate vibrational frequencies 
yields similar looking formulae. 
However, that method cannot be used for the exact fn sampling of general 
operators, only gives the fn response with respect to a parameter such as the 
distance between
 atoms, and applies specifically to DMC with fully retained weights.

\bibitem{casino} R. J. Needs, M. Towler, N. Drummond, and P. Kent, {\it CASINO
version 1.7 User manual} (University of Cambridge, 2004).
\bibitem{pines} D. Pines and P. Noziers, The Theory of Quantum Liquids 
(Benjamin, New York, 1966).
\bibitem{fxcref} B. Farid, V. Heine, G. E. Engel, and I. J. Robertson, Phys.
Rev. B {\bf 48}, 11602 (1993).



\end{thebibliography}
\end{document}